\documentclass{elsart} 
\journal{Physics Letters B}

\begin{document}
\begin{frontmatter}

\title{A Lagrangian description of interacting dark energy} 
\author{Nikodem J. Pop\l awski}
\ead{nipoplaw@indiana.edu}
\address{Department of Physics, Indiana University,
727 East Third Street, Bloomington, IN 47405, USA}

\begin{abstract}
We propose a relativistically covariant model of interacting dark energy
based on the principle of least action.
The cosmological term $\Lambda$ in the gravitational Lagrangian is a function of
the trace of the energy--momentum tensor $T$.
We find that the $\Lambda(T)$ gravity is more general than the Palatini $f(R)$ gravity,
and reduces to the latter if we neglect the pressure of matter.
We show that recent cosmological data favor a variable cosmological constant
and are consistent with the $\Lambda(T)$ gravity, without knowing
the specific function $\Lambda(T)$.
\end{abstract}

\end{frontmatter}

\section{Introduction}

The most accepted explanation of the observed cosmic
acceleration~\cite{univ} is that the universe is dominated by an exotic
component with large negative pressure, called dark energy~\cite{DE}.
The simplest way of introducing dark energy into general relativity is
to add a cosmological constant $\Lambda$ to the
Riemann curvature scalar $R$ in the Lagrangian:
\begin{equation}
S=S_m-\frac{1}{2\kappa c}\int d^4 x\sqrt{-g}[R+2\Lambda],
\label{action1}
\end{equation}
which leads to Friedmann and Lema\^{i}tre cosmological models~\cite{FL}. 
The $\Lambda$ term is spatially uniform and time independent (explicitly),
as required by the principle of general covariance, and can be viewed as a
representation of a relativistic ideal fluid obeying the equation of state
$p=-\epsilon$.
Moreover, this fluid obeys the equation of continuity that does not depend on
matter energy density.
Therefore, such a form of dark energy is said to be non-interacting.

An interaction between ordinary matter and dark energy can be introduced in
the form of a time dependent cosmological constant $\Lambda(t)$~\cite{var}.
A decreasing $\Lambda$ gives a positive rate of the entropy production
and can be responsible for the observed large entropy of the universe.
The cosmological constant can depend on the cosmic time through the scale
factor $a$ or the Hubble parameter $H$, and
dimensional arguments lead to $\Lambda\propto a^{-2}$ or
$\Lambda\propto H^2$, respectively~\cite{R,H}.
This dependence can, of course, have a different form and involve other
quantities such as temperature.
Phenomenological models with a variable cosmological constant are listed
and reviewed in~\cite{OC}.
However, they all lack a Lagrangian description, i.e. the interaction
between matter and dark energy cannot be derived in these models from
the principle of least action in a relativistically covariant form.

To preserve the covariance, a variable cosmological constant must
depend only on relativistic invariants.
A possible choice is $\Lambda$ as a function of $R$.
Such models are referred to as $f(R)$ gravity and can explain current
cosmic acceleration, e.g., for $\Lambda(R)\propto R^{-1}$~\cite{acc1,acc2}.
In these models, a Legendre--Helmholtz transformation of the Lagrangian or
a conformal transformation of the metric bring the gravitational field
equations of $f(R)$ gravity into the form of the Einstein equations of
general relativity with an additional scalar field~\cite{eq}.
A similar equivalence can be proven for a Lagrangian depending on
invariants composed of the Ricci tensor $R_{\mu\nu}$~\cite{JK}.

In this Letter, we propose another choice in which the cosmological constant
is a function of the trace of the energy--momentum tensor,
and dub such a model $\Lambda(T)\,gravity$.
This choice has an advantage over $f(R)$ and $f(R_{\mu\nu})$ gravity theories
because we do not need to transform (conformally) from one metric to another, and there is no question about which metric is physical since we use one and
the same metric tensor~\cite{eq}.
In Sec.~2 we derive the field equations in $\Lambda(T)$ gravity and apply
them to the Robertson--Walker metric which describes a homogeneous and isotropic universe.
In Sec.~3 we relate our model to $f(R)$ gravity and some phenomenological models
of interacting dark energy.
In Sec.~4 we show that the cosmological data favor a variable cosmological constant.
The results are summarized in Sec.~5.

\section{The field equations and cosmology}

The $\Lambda(T)$ gravity action is given by
\begin{equation}
S=S_m-\frac{1}{2\kappa c}\int d^4 x\sqrt{-g}[R+2\Lambda(T)],
\label{action2}
\end{equation} 
and the dynamical energy--momentum tensor $T_{\mu\nu}$ is generated from the action for matter $S_m$:
\begin{equation}
\delta S_m=\frac{1}{2c}\int d^4x\sqrt{-g} T_{\mu\nu}\delta g^{\mu\nu}.
\label{emt}
\end{equation}
From the principle of least action, the variation of~(\ref{action2}) with respect to the metric tensor $g_{\mu\nu}$ equals zero,
\begin{eqnarray}
& & \delta S_m-\frac{1}{2\kappa c}\int d^4x\sqrt{-g}\biggl(R_{\mu\nu}-\frac{1}{2}Rg_{\mu\nu}+2\Lambda'(T)T_{\mu\nu}-\Lambda(T)g_{\mu\nu}\biggr)\delta g^{\mu\nu} \nonumber \\
& & -\frac{1}{\kappa c}\int d^4x\sqrt{-g}\Lambda'(T)g^{\mu\nu}\delta T_{\mu\nu}=0,
\label{var1}
\end{eqnarray}
where the prime denotes the differentiation with respect to $T$.
For the variation of $T_{\mu\nu}$ we write a general expression,
\begin{equation}
\delta T_{\mu\nu}=N_{\mu\nu\rho\sigma}\delta g^{\rho\sigma}=-N_{\mu\nu}^{\phantom{\mu\nu}\rho\sigma}\delta g_{\rho\sigma},
\label{var2}
\end{equation}
where the tensor $N_{\mu\nu\rho\sigma}$ depends on the kind of matter described by $T_{\mu\nu}$.
Eqs.~(\ref{var1}) and~(\ref{var2}) yield the Einstein field equations,\footnote{
The cosmological term $\Lambda(T)$ modifies the gravitational
interaction between matter and curvature, replacing $\kappa$ by the
running gravitational coupling parameter which is a function of the tensors
$T_{\mu\nu}$ and $N^{\rho}_{\phantom{\rho}\rho\mu\nu}$.}
\begin{equation}
R_{\mu\nu}-\frac{1}{2}Rg_{\mu\nu}=[\kappa-2\Lambda'(T)]T_{\mu\nu}+\Lambda(T)g_{\mu\nu}-2\Lambda'(T)N^{\rho}_{\phantom{\rho}\rho\mu\nu}.
\label{Ein1}
\end{equation}

For a massless scalar field, $T_{\mu\nu}=\frac{1}{2}\partial_\mu\phi\partial_\nu\phi-\frac{1}{4}g_{\mu\nu}\partial_\rho\phi\partial^\rho\phi$, we find
\begin{eqnarray}
& & N_{\mu\nu\rho\sigma}=\frac{1}{4}g_{\rho(\mu}g_{\nu)\sigma}\partial_\lambda\phi\partial^\lambda\phi-\frac{1}{4}g_{\mu\nu}\partial_\rho\phi\partial_\sigma\phi,
\label{sc1} \\
& & N^{\rho}_{\phantom{\rho}\rho\mu\nu}=\frac{1}{4}g_{\mu\nu}\partial_\rho\phi\partial^\rho\phi-\partial_\mu\phi\partial_\nu\phi=-2T_{\mu\nu}+\frac{1}{2}Tg_{\mu\nu},
\label{sc2}
\end{eqnarray}
where the brackets denote symmetrization.
In this case, Eq.~(\ref{Ein1}) becomes
\begin{equation}
R_{\mu\nu}-\frac{1}{2}Rg_{\mu\nu}=[\kappa+2\Lambda'(T)]T_{\mu\nu}+[\Lambda(T)-T\Lambda'(T)]g_{\mu\nu}.
\label{sc3}
\end{equation}
For the electromagnetic field, $T_{\mu\nu}=\frac{1}{4}g_{\mu\nu}F_{\rho\sigma}F^{\rho\sigma}-F_{\mu\rho}F_{\nu}^{\phantom{\nu}\rho}$, we obtain
\begin{eqnarray}
& & N_{\mu\nu\rho\sigma}=F_{\rho(\mu}F_{\nu)\sigma}+\frac{1}{2}g_{\mu\nu}F_{\rho\lambda}F_{\sigma}^{\phantom{\sigma}\lambda}-\frac{1}{4}g_{\rho(\mu}g_{\nu)\sigma}F_{\kappa\lambda}F^{\kappa\lambda},
\label{em1} \\
& & N^{\rho}_{\phantom{\rho}\rho\mu\nu}=F_{\mu\rho}F_{\nu}^{\phantom{\nu}\rho}-\frac{1}{4}g_{\mu\nu}F_{\rho\sigma}F^{\rho\sigma}=-T_{\mu\nu}.
\label{em2}
\end{eqnarray}
In this case, Eq.~(\ref{Ein1}) reads simply
\begin{equation}
R_{\mu\nu}-\frac{1}{2}Rg_{\mu\nu}=\kappa T_{\mu\nu}+\Lambda(T)g_{\mu\nu}.
\label{em3}
\end{equation}
The term $\Lambda(T)$ does not affect the electromagnetic field since $T=0$.
For a perfect fluid, $T_{\mu\nu}=(\epsilon+p)u_{\mu\nu}-pg_{\mu\nu}$, we derive
\begin{eqnarray}
& & N_{\mu\nu\rho\sigma}=pg_{\rho(\mu}g_{\nu)\sigma}-(\epsilon+p)(g_{\rho(\mu}u_{\nu)}u_\sigma+g_{\sigma(\mu}u_{\nu)}u_\rho),
\label{fl1} \\
& & N^{\rho}_{\phantom{\rho}\rho\mu\nu}=pg_{\mu\nu}-2(\epsilon+p)u_\mu u_\nu=-2T_{\mu\nu}-pg_{\mu\nu}.
\label{fl2}
\end{eqnarray}
Unlike for the previous cases, the tensor $N^{\rho}_{\phantom{\rho}\rho\mu\nu}$ cannot be now expressed as a linear combination of $T_{\mu\nu}$ and $Tg_{\mu\nu}$.\footnote{
Contracting Eq.~(\ref{Ein1}) with the metric tensor gives $-R=[\kappa-2\Lambda'(T)]T+4\Lambda(T)-2\Lambda'(T)N^{\rho\phantom{\rho\sigma}\sigma}_{\phantom{\rho}\rho\sigma}$.
The impossibility of expressing the tensor $N^{\rho}_{\phantom{\rho}\rho\mu\nu}$ as a function of $T_{\mu\nu}$ and $Tg_{\mu\nu}$ means that the scalar $N^{\rho\phantom{\rho\sigma}\sigma}_{\phantom{\rho}\rho\sigma}$ is not a function of $T$ only.
For some special cases, such as a massless scalar field or a pressureless dust, the quantity $N^{\rho\phantom{\rho\sigma}\sigma}_{\phantom{\rho}\rho\sigma}$ is a definite function of $T$ and so is $R$.
Therefore, if we know the function $\Lambda(T)$, we obtain an algebraic equation for $T=T(R)$ and
the $\Lambda(T)$ gravity reduces to the Palatini $f(R)$ gravity formulated in the Einstein frame (see the end of Sec.~3)~\cite{eq,Niko}.}
The field equations~(\ref{Ein1}) give
\begin{equation}
R_\mu^\nu-\frac{1}{2}R\delta_\mu^\nu=[\kappa+2\Lambda'(T)](\epsilon+p)u_\mu u^\nu+[\Lambda(T)-\kappa p]\delta_\mu^\nu.
\label{fl3}
\end{equation}

The Bianchi identity requires the right-hand side of~(\ref{Ein1}) to be
covariantly conserved, which leads to
\begin{eqnarray}
& & T_{\mu\nu}^{\phantom{\mu\nu};\nu}=\{2\Lambda''(T)T^{,\nu}T_{\mu\nu}-\Lambda'(T)T_{,\mu}+2\Lambda''(T)T^{,\nu}N^\rho_{\phantom{\rho}\rho\mu\nu} \nonumber \\
& & +2\Lambda'(T)N^{\rho\phantom{\rho\mu\nu};\nu}_{\phantom{\rho}\rho\mu\nu}\}[\kappa-2\Lambda'(T)]^{-1}.
\label{Bian}
\end{eqnarray}
Eq.~(\ref{Bian}) means that the energy--momentum tensor is not,
in general, covariantly conserved.
We can write the field equations~(\ref{Ein1}) as
\begin{equation}
R_{\mu\nu}-\frac{1}{2}Rg_{\mu\nu}=\kappa(T_{\mu\nu}^m+T_{\mu\nu}^{\Lambda}),
\label{Ein2}
\end{equation}
where $T_{\mu\nu}^m=T_{\mu\nu}$, which defines the dark energy--momentum tensor,\footnote{
Such defined dark energy--momentum tensor depends on the tensor $T_{\mu\nu}$. Therefore, the former represents both ``pure'' dark energy (vacuum) and the interaction between vacuum and matter.} 
\begin{equation}
T_{\mu\nu}^{\Lambda}=\frac{1}{\kappa}[\Lambda(T)g_{\mu\nu}-2\Lambda'(T)(T_{\mu\nu}+N^{\rho}_{\phantom{\rho}\rho\mu\nu})].
\label{dark1}
\end{equation}
From Eq.~(\ref{Ein2}) it follows that matter and dark energy
form together a system that has a conserved four-momentum.
Therefore, we can speak of an interaction between matter and dark
energy that together form a closed system.
For a perfect fluid, we can write Eq.~(\ref{dark1}) as
\begin{equation}
T_{\mu\nu}^\Lambda=(\epsilon_\Lambda+p_\Lambda)u_\mu u_\nu-p_\Lambda g_{\mu\nu},
\label{dark2}
\end{equation}
from which we obtain the expressions for the dark
energy density $\epsilon_\Lambda$ and pressure $p_\Lambda$:
\begin{eqnarray}
& & \epsilon_\Lambda=\frac{2\Lambda'(T)}{\kappa}(\epsilon_m+p_m)+\frac{\Lambda(T)}{\kappa}, \label{de1} \\
& & p_\Lambda=-\frac{\Lambda(T)}{\kappa}, \label{de2}
\end{eqnarray}
where $T=\epsilon-3p$.

We now apply Eq.~(\ref{fl3}) to a flat~\cite{Gold}, homogeneous and isotropic universe filled with a perfect fluid, described by the Robertson--Walker metric,
\begin{equation}
ds^2=c^2dt^2-a^2(t)(dr^2+r^2d\Omega^2).
\label{RW}
\end{equation}
The first Friedmann equation gives the Hubble parameter,
\begin{equation}
H=\frac{\dot{a}}{a}=c\sqrt{\frac{(\kappa+2\Lambda')\epsilon_m+2\Lambda'p_m+\Lambda}{3}},
\label{Hub1}
\end{equation}
where $\Lambda=\Lambda(T)$ and $\Lambda'=\Lambda'(T)$.
The dot denotes the differentiation with respect to the cosmic time $t$.
The second Friedmann equation reads
\begin{equation}
\frac{\dot{a}^2+2a\ddot{a}}{c^2 a^2}=-\kappa p_m+\Lambda,
\label{dec1}
\end{equation}
which yields the deceleration parameter,
\begin{equation}
q=-\frac{a\ddot{a}}{\dot{a}^2}=\frac{(\kappa+2\Lambda')\epsilon_m+(3\kappa+2\Lambda')p_m-2\Lambda}{2[(\kappa+2\Lambda')\epsilon_m+2\Lambda'p_m+\Lambda]}.
\label{dec2}
\end{equation}
The density ratio parameter $\Omega_\Lambda=\epsilon_\Lambda/\epsilon_c$,
where $\epsilon_c=3H^2/(\kappa c^2)$ is the critical energy density,
is given by\footnote{
The $\epsilon_\Lambda$ defined in Eq.~(\ref{de1}) satisfies $\epsilon_m+\epsilon_\Lambda=\epsilon_c$.
Therefore, $\Omega=\Omega_m+\Omega_\Lambda=(\epsilon_m+\epsilon_\Lambda)/\epsilon_c=1$, as in general-relativistic cosmology for a flat universe.}
\begin{equation}
\Omega_\Lambda=\frac{2\Lambda'(\epsilon_m+p_m)+\Lambda}{(\kappa+2\Lambda')\epsilon_m+2\Lambda'p_m+\Lambda}.
\label{Omega1}
\end{equation}

We restrict our attention to a universe filled with dust, for which $p_m=0$ and
$T=\epsilon$.
Hereinafter, the prime denotes the differentiation with respect to the energy density $\epsilon$, and we write $\epsilon$ instead of $\epsilon_m$.
Eqs.~(\ref{Hub1}) and~(\ref{dec2}) become, respectively,\footnote{
In the case of dust, Eq.~(\ref{fl3}) can be written as $R_{\mu\nu}-\frac{1}{2}Rg_{\mu\nu}=\kappa_r T_{\mu\nu}+\Lambda(T)g_{\mu\nu}$, where the running gravitational coupling is given by $\kappa_r=\kappa+2\Lambda'(T)$.}
\begin{eqnarray}
& & H=c\sqrt{\frac{(\kappa+2\Lambda')\epsilon+\Lambda}{3}},
\label{Hub2} \\
& & q=\frac{(\kappa+2\Lambda')\epsilon-2\Lambda}{2[(\kappa+2\Lambda')\epsilon+\Lambda]}
\label{dec3},
\end{eqnarray}
while Eq.~(\ref{Omega1}) reduces to
\begin{equation}
\Omega_\Lambda=\frac{2\Lambda'\epsilon+\Lambda}{(\kappa+2\Lambda')\epsilon+\Lambda}.
\label{Omega2}
\end{equation}
When the universe is dominated by matter, $q\sim1/2$ and $\Omega_\Lambda\sim0$, and when it is
dominated by dark energy, $q\sim-1$ and $\Omega_\Lambda\sim1$.
The deceleration-to-acceleration transition ($q=0$) takes place when
$\epsilon=\epsilon_t$, where $\epsilon_t$ is determined by
\begin{equation}
[\kappa+2\Lambda'(\epsilon_t)]\epsilon_t-2\Lambda(\epsilon_t)=0.
\label{trans}
\end{equation}

Applying the Bianchi identity to Eq.~(\ref{Ein2}) leads to the equation of continuity,
\begin{equation}
(\epsilon+\epsilon_\Lambda)^{\cdot}+3H(\epsilon+\epsilon_\Lambda+p_\Lambda)=0,
\label{cont1}
\end{equation}
which gives the time dependence of the matter energy density,
\begin{equation}
\dot{\epsilon}=-3H\epsilon\,\frac{\kappa+2\Lambda'}{\kappa+3\Lambda'+2\Lambda''\epsilon}.
\label{cont2}
\end{equation}
Using $a=a_0(1+z)^{-1}$ and integrating~(\ref{cont2}) gives the relation
between the energy density and redshift,
\begin{equation}
\mbox{ln}(1+z)=\frac{1}{3}\int_{\epsilon_0}^{\epsilon}\frac{d\epsilon}{\epsilon}\,\frac{\kappa+3\Lambda'+2\Lambda''\epsilon}{\kappa+2\Lambda'},
\label{cont3}
\end{equation}
where $\epsilon_0$ is the present energy density.
Combining~(\ref{Hub2}) with~(\ref{cont3}) yields the relation $H=H(z)$ which
is of observational interest.
If $\Lambda=\mbox{const}$, we reproduce the $\Lambda CDM$ relation $H^2(z)=\frac{c^2}{3}[\kappa\epsilon_0(1+z)^3+\Lambda]$.

We can write the equation of continuity~(\ref{cont1}) as
\begin{eqnarray}
& & \dot{\epsilon}+3H\epsilon=Q, \label{con1} \\
& & \dot{\epsilon}_\Lambda+3H(\epsilon_\Lambda+p_\Lambda)=-Q, \label{con2}
\end{eqnarray}
where $Q$ is the rate of the energy transfer from dark energy to ordinary
matter,
\begin{equation}
Q=3H\epsilon\,\frac{\Lambda'+2\Lambda''\epsilon}{\kappa+3\Lambda'+2\Lambda''\epsilon}.
\label{Q}
\end{equation}
The relative transfer rate $\Gamma$ equals $Q/\epsilon_\Lambda$,
\begin{equation}
\Gamma=3\kappa H\epsilon\,\frac{\Lambda'+2\Lambda''\epsilon}{(\kappa+3\Lambda'+2\Lambda''\epsilon)(2\Lambda'\epsilon+\Lambda)}.
\label{Gamma}
\end{equation}
If $\Lambda=0$ then both $Q$ and $\epsilon_\Lambda$ vanish and $\Gamma$ is
indeterminate.
The non-dimensional interaction rate $\tau=\Gamma/H$ is given by
\begin{equation}
\tau=\frac{3\kappa\epsilon(\Lambda'+2\Lambda''\epsilon)}{(\kappa+3\Lambda'+2\Lambda''\epsilon)(2\Lambda'\epsilon+\Lambda)}.
\label{ratio}
\end{equation}
The first and second law of thermodynamics for an isotropic and homogeneous
universe can be written as $(\epsilon a^3)^{\cdot}+p(a^3)^{\cdot}=T\dot{S}$, which with Eq.~(\ref{con1}) yields
\begin{equation}
TdS=a^3 Qdt.
\label{the}
\end{equation}
The quantity $Q$ is related to the production of entropy in a growing universe,
and thus must be positive.

\section{$\Lambda(T)$ gravity and phenomenological models}

If $\Lambda=\mbox{const}$, we return to the $\Lambda CDM$ cosmology.
If $\Lambda$ is proportional to $T=\epsilon$, $\Lambda=n\kappa\epsilon$,
where $n\neq0$ is a non-dimensional constant, we reproduce the model with
$\Lambda\propto H^2$~\cite{H}:
\begin{equation}
\Lambda=\frac{3nH^2}{(1+3n)c^2},\,\,\,\,\tau=\frac{1}{1+3n}, 
\label{lin}
\end{equation}
where the Hubble parameter is given by
\begin{equation}
H=c\sqrt{\kappa\epsilon\biggl(n+\frac{1}{3}\biggr)}.
\label{Hub3}
\end{equation}
We have a static case $H=0$ for $n=-\frac{1}{3}$, and no solution if $n<-\frac{1}{3}$.
From non-negativity of $Q$ and $\epsilon$ it follows that $n\geq0$.
Eq.~(\ref{cont2}) and the definition of $H$ yield the law
for scaling of the energy density,
\begin{equation}
\epsilon\propto a^{-3\bigl(\frac{1+2n}{1+3n}\bigr)}.
\label{scal}
\end{equation}
There is no deceleration-to-acceleration transition, however, since
\begin{equation}
q=\frac{1}{2(1+3n)}=\mbox{const}.
\label{dec4}
\end{equation}

To explain the observed transition, we need to break the proportionality
$\Lambda\propto\epsilon$.
The simplest choice is $\Lambda=n\kappa\epsilon+\Lambda_0$, where
$\Lambda_0$ is a constant\footnote{
This form can be viewed as the linear approximation of the real function $\Lambda(\epsilon)$.}.
The scaling law~(\ref{scal}) does not change, while the deceleration
parameter becomes
\begin{equation}
q=\frac{\kappa\epsilon-2\Lambda_0}{2[(1+3n)\kappa\epsilon+\Lambda_0]}.
\label{decel}
\end{equation}
The transition occurs at
\begin{equation}
\epsilon_t=\frac{2\Lambda_0}{\kappa},
\label{transit1}
\end{equation}
from which we obtain, using Eq.~(\ref{cont3}), the following relation:
\begin{equation}
z_t=\biggl(\frac{2\Lambda_0}{\epsilon_0\kappa}\biggr)^{\bigl(\frac{1+3n}{3(1+2n)}\bigr)}-1.
\label{transit2}
\end{equation}
By means of Eq.~(\ref{Omega2}), we derive
\begin{equation}
z_t=\biggl[2\frac{(1+3n)\Omega_{\Lambda,0}-3n}{1-\Omega_{\Lambda,0}}\biggr]^{\bigl(\frac{1+3n}{3(1+2n)}\bigr)}-1,
\label{transit3}
\end{equation}
where the subscript $0$ refers to the present time.
The Hubble parameter as a function of the redshift can be found using Eqs.~(\ref{Hub2}) and~(\ref{scal}):
\begin{equation}
H(z)=\frac{c}{\sqrt{3}}\biggl[(1+3n)\kappa\epsilon_0(1+z)^{\bigl(\frac{3(1+2n)}{1+3n}\bigr)}+\Lambda_0\biggr]^{1/2}.
\label{last1}
\end{equation}
By means of Eq.~(\ref{de1}) and the definitions of $\Omega_{m,0}$ and $\Omega_{\Lambda,0}$, we bring Eq.~(\ref{last1}) to the standard form,
\begin{equation}
H(z)=H_0\biggl[(1+3n)\Omega_{m,0}(1+z)^{\bigl(\frac{3(1+2n)}{1+3n}\bigr)}-3n\Omega_{m,0}+\Omega_{\Lambda,0}\biggr]^{1/2}.
\label{last2}
\end{equation}
For $n=0$, Eqs.~(\ref{transit3}) and~(\ref{last2}) reduce to their $\Lambda CDM$ expressions.

The relation $\Lambda\propto H^2$ holds for a larger class of functions
$\Lambda(\epsilon)$ obeying
\begin{equation}
(\kappa+2\Lambda')\epsilon+(1-k^{-1})\Lambda=0,
\label{LH1}
\end{equation}
where the constant of proportionality $k$ is defined such that $\Lambda=3kH^2/c^2$.\footnote{
The constants $n$ and $k$ are related by $k=\frac{n}{1+3n}$.}
The solution of~(\ref{LH1}) is given by
\begin{equation}
\Lambda=\frac{k\kappa\epsilon}{1-3k}+C\epsilon^{\frac{1-k}{2k}}.
\label{LH2}
\end{equation}
Hereinafter, $C$ denotes a constant.
The singular case $\Lambda=H^2/c^2$ can be obtained from
$\Lambda=-\frac{\kappa}{2}\epsilon\,\mbox{ln}\epsilon+C\epsilon$.
The general condition for the constancy of $\tau$ leads to a more complicated
differential equation for $\Lambda$.

Another phenomenological condition $\Lambda\propto a^{-m}$, where $m=\mbox{const}$, can be found by combining~(\ref{cont2}) with the identity
$\dot{\epsilon}=(d\epsilon/d\Lambda)(d\Lambda/da)Ha$.
We find the second-order differential equation for $\Lambda(\epsilon)$ which is equivalent to this condition:
\begin{equation}
m\Lambda(\kappa+3\Lambda'+2\Lambda''\epsilon)-3\epsilon\Lambda'(\kappa+2\Lambda')=0.
\label{La}
\end{equation}
This equation has a solution linear in $\epsilon$, $\Lambda=\frac{3-m}{3(m-2)}\kappa\epsilon$,
except for the dimensionally favored case $m=2$~\cite{R}.

The condition for no interaction between matter and dark energy follows
from~(\ref{cont2}) or~(\ref{Q}):
\begin{equation}
\Lambda'+2\Lambda''\epsilon=0.
\label{noint1}
\end{equation}
The solution of this equation is given by
\begin{equation}
\Lambda=C\epsilon^{1/2}+\Lambda_0.
\label{noint2}
\end{equation}
In this case, the deceleration parameter vanishes when
\begin{equation}
\kappa\epsilon-C\sqrt{\epsilon}-2\Lambda_0=0,
\label{noint3}
\end{equation}
which has one physical solution for $\epsilon$. 

Finally, we examine the case in which $\Lambda$ is a linear function of $q$.
The relation $\Lambda\propto q$ gives a nonlinear differential equation
for the function $\Lambda(\epsilon)$.
The Palatini $f(R)$ gravity in the Einstein conformal frame predicts
\begin{equation}
\Lambda=\frac{H^2}{c^2}(1-2q),
\label{Hq}
\end{equation}
which is linear in $q$ but also includes the dimensional factor $H^2$~\cite{IDE}.
Using Eqs.~(\ref{Hub2}) and~(\ref{dec3}) we can show that the same relation~(\ref{Hq}) is satisfied in the $\Lambda(T)$ gravity. 
Therefore, we conclude that the $\Lambda(T)$ gravity and the Palatini $f(R)$ gravity are
cosmologically equivalent, although this is true only for matter without pressure.
For more general cases, including matter described by both $\epsilon$ and $p$, the $\Lambda(T)$ gravity is broader than the Palatini $f(R)$ gravity.
In the latter, $R$ is solely a function of $T$~\cite{acc1}, which is generally not true for the $\Lambda(T)$ gravity.\footnote{
See the second footnote in Sec.~2.}

\section{Observational constraints}

Recent WMAP cosmological data~\cite{WMAP} indicate that matter in the universe scales almost like dust, $\epsilon a^3\sim\mbox{const}$.
If we define $\varepsilon$ as the deviation from the factor 3 in the scaling law,
\begin{equation}
\epsilon\sim a^{-(3-\varepsilon)},
\label{dev1}
\end{equation}
then the data require $0\leq\varepsilon\leq0.1$~\cite{wm}.
Assuming $\Lambda=n\kappa\epsilon+\Lambda_0$ gives, using
Eq.~(\ref{scal}) and~(\ref{dev1}), the equivalent condition
\begin{equation}
0\leq n\leq0.037.
\label{dev2}
\end{equation}
The current value of the dark energy density parameter is $\Omega_{\Lambda,0}=0.71^{+0.03}_{-0.05}$~\cite{Gold}, which in the $\Lambda CDM$ model gives $z_t=0.70^{+0.08}_{-0.13}$.
This value of $z_t$ slightly overlaps with the observed $z_t=0.46\pm0.13$~\cite{Gold}.
For the largest $n$ compatible with observations, $n=0.037$, we use Eq.~(\ref{transit3}) to find
\begin{equation}
z_t=0.70^{+0.10}_{-0.14}.
\label{dev3}
\end{equation}
This value slightly overlaps with the observed $z_t$ as well.
Consequently, the $\Lambda(\epsilon)=n\kappa\epsilon+\Lambda_0$ model is consistent
with astronomical observations, provided $n\ll1$, which means that the interaction
between dark energy and ordinary matter is relatively weak~\cite{IDE}.

We now proceed to show that the cosmological data favor a variable cosmological constant.
If we define
\begin{equation}
\alpha=\frac{\Lambda_0}{\kappa\epsilon_0},\,\,\,\,\beta=\frac{\Lambda'_0}{\kappa},\,\,\,\gamma=\frac{\Lambda''_0\epsilon_0}{\kappa},
\label{abg}
\end{equation}
then we can express $\Omega_{m,0}$ and $q_0$ as
\begin{equation}
\Omega_{m,0}=(1+\alpha+2\beta)^{-1},\,\,\,\,q_0=\frac{1-2\alpha+2\beta}{2(1+\alpha+2\beta)}.
\label{zero1}
\end{equation}
Reversing these equations yields
\begin{equation}
\alpha=\frac{1-2q_0}{3\Omega_{m,0}},\,\,\,\,\beta=\frac{2+2q_0-3\Omega_{m,0}}{6\Omega_{m,0}}.
\label{zero2}
\end{equation}
In the $\Lambda CDM$ model, $\beta=0$ and $q_0=3\Omega_{m,0}/2-1$.

The values for the present matter density parameter and
deceleration parameter from the SNIa gold sample data are
$\Omega_{m,0}=0.29^{+0.05}_{-0.03}$ and
$q_0=-0.74\pm0.18$, respectively~\cite{Gold}.
They give
\begin{eqnarray}
& & \alpha=2.85^{+0.79}_{-0.77}, \label{obs1} \\
& & \beta=-0.20^{+0.26}_{-0.22}. \label{obs2}
\end{eqnarray}
The result of~(\ref{obs2}) favors a variable cosmological constant
($\beta\neq0$), although does not exclude a possibility of $\Lambda$ being
a constant ($\beta=0$).\footnote{
In the case $\Lambda(\epsilon)=n\kappa\epsilon+\Lambda_0$, we have $\beta=n$.
Since $n$ cannot be negative, we are left with the condition $n\leq0.06$
which is on the order of the previously found inequality~(\ref{dev2}).}
The redshift of the deceleration-to-acceleration transition can be found
from equations~(\ref{trans}) and~(\ref{cont3}), and depends on a
particular form of the function $\Lambda(\epsilon)$.

Another non-dimensional parameter measured by cosmologists
is the current value for the redshift derivative of the deceleration
parameter, $(dq/dz)_0$.
We use the identity $dq/dz=(dq/d\epsilon)(dz/d\epsilon)^{-1}$,
Eq.~(\ref{cont3}) written as
\begin{equation}
\frac{dz}{d\epsilon}\bigg|_0=\frac{1+3\beta+2\gamma}{3\epsilon_0(1+2\beta)},
\label{slo1}
\end{equation}
and the derivative of Eq.~(\ref{dec3}) with respect to $\epsilon$,
to obtain
\begin{equation}
\frac{dq}{dz}\bigg|_0=\frac{3(1+2\beta)(2\alpha-3\beta+3\alpha\beta-6\beta^2+4\alpha\gamma)}{2(1+3\beta+2\gamma)(1+\alpha+2\beta)^2}.
\label{slo2}
\end{equation}
Let us assume $\beta=0$.
In this case, Eq.~(\ref{slo2}) becomes
\begin{equation}
\frac{dq}{dz}\bigg|_0=\frac{3\alpha}{(1+\alpha)^2},
\label{slo3}
\end{equation}
and is independent of $\gamma$.
Substituting here $\alpha$ from~(\ref{obs1}) yields
\begin{equation}
\frac{dq}{dz}\bigg|_0=0.57^{+0.09}_{-0.07},
\label{slo4}
\end{equation}
which does not overlap with the observed $1.59\pm0.63$~\cite{Gold}.
Therefore, the SNIa gold sample data for $(dq/dz)_0$ support
$\beta\neq0$.
We emphasize that the above derivation is general, since we did not use the
specific function $\Lambda(\epsilon)$.

\section{Summary}

The presented $\Lambda(T)$ gravity is a relativistically covariant model
that predicts a variable cosmological constant.
It derives the field equations and the interaction between matter and dark energy
from the principle of least action, generating this interaction
from the cosmological term $\Lambda$ which is a function of
the trace of the energy--momentum tensor $T$.
We showed that the $\Lambda(T)$ gravity and the Palatini $f(R)$ gravity give
the same dynamics of the universe filled with dust, i.e. they are
cosmologically equivalent if $T=\epsilon$.
For more general cases, the $\Lambda(T)$ gravity is broader.
We also showed that cosmological data favor the cosmological
constant which varies with time.
Further astronomical observations should provide insight into the form of
the function $\Lambda(T)$, describing how matter interacts with dark energy
and how dark energy affects the gravitational coupling between matter
and spacetime.

\section*{Acknowledgment}

The author would like to thank Tomi Koivisto for valuable comments.


\begin{thebibliography}{25}
\bibitem{univ} A. G. Riess {\it et al.}, Astron. J. {\bf 116}, 1009 (1998);
S. Perlmutter {\it et al.}, Astrophys. J. {\bf 517}, 565 (1999);
N. W. Halverson {\it et al.}, Astrophys. J. {\bf 568}, 38 (2002); 
C. B. Netterfield {\it et al.}, Astrophys. J. {\bf 571}, 604 (2002);
C. L. Bennett {\it et al.}, Astrophys. J. Suppl. Ser. {\bf 148}, 1 (2003).
\bibitem{DE} P. J. E. Peebles and B. Ratra, Rev. Mod. Phys. {\bf 75}, 559 (2003);
T. Padmanabhan, Phys. Rept. {\bf 380}, 235 (2003);
E. J. Copeland, M. Sami, and S. Tsujikawa, hep-th/0603057 (2006).
\bibitem{FL} J. E. Felten and R. Isaacman, Rev. Mod. Phys. {\bf 58}, 689 (1986).
\bibitem{var} M. Bronstein, Phys. Z. Sowjetunion {\bf 3}, 73 (1933);
M. S. Berman, Phys. Rev. D {\bf 43}, 1075 (1991).
\bibitem{R} M. \"{O}zer and M. O. Taha, Phys. Lett. B {\bf 171}, 363 (1986);
W. Chen and Y. S. Wu, Phys. Rev. D {\bf 41}, 695 (1990).
\bibitem{H} J. C. Carvalho, J. A. S. Lima, and I. Waga, Phys. Rev. D {\bf 46}, 2404 (1992);
J. A. S. Lima and J. C. Carvalho, Gen. Relativ. Gravit. {\bf 26}, 909 (1994).
\bibitem{OC} J. M. Overduin and F. I. Cooperstock, Phys. Rev. D {\bf 58}, 043506 (1998).
\bibitem{acc1} D. N. Vollick, Phys. Rev. D {\bf 68}, 063510 (2003).
\bibitem{acc2} S. M. Carroll, V. Duvvuri, M. Trodden, and M. S. Turner, Phys. Rev. D {\bf 70}, 043528 (2004).
\bibitem{eq} G. Magnano, gr-qc/9511027 (1995).
\bibitem{JK} A. Jakubiec and J. Kijowski, Phys. Rev. D {\bf 37}, 1406 (1988).
\bibitem{Niko} N. J. Pop\l awski, Class. Quantum Grav. {\bf 23}, 2011 {\it ibid} 4819 (2006).
\bibitem{Gold} A. G. Riess {\it et al.}, Astrophys. J. {\bf 607}, 665 (2004).
\bibitem{IDE} N. J. Pop\l awski, gr-qc/0607124 (2006).
\bibitem{WMAP} D. N. Spergel {\it et al.}, Astrophys. J. Suppl. Ser. {\bf 148}, 175 (2003).
\bibitem{wm} P. Wang and X. H. Meng, Class. Quantum Grav. {\bf 22}, 283 (2005).
\end{thebibliography}
\end{document}